\documentclass[conference]{IEEEtran}
\IEEEoverridecommandlockouts
\usepackage{cite}
\usepackage{amsmath,amssymb,amsfonts}
\usepackage{algorithmic}
\usepackage{graphicx}
\usepackage{textcomp}
\usepackage{xcolor}
\def\BibTeX{{\rm B\kern-.05em{\sc i\kern-.025em b}\kern-.08em
    T\kern-.1667em\lower.7ex\hbox{E}\kern-.125emX}}

\usepackage{comment}

\usepackage[detect-all=true]{siunitx}
\DeclareSIUnit\decibelm{dBm}
\DeclareSIUnit\packets{packets}
\DeclareSIUnit\pct{percentile}
\AtBeginDocument{
	\DeclareSIUnit\bit{b}
	\DeclareSIUnit\bitpersec{bps}

	\newcolumntype{C}[1]{>{\centering\let\newline\\\arraybackslash\hspace{0pt}}m{#1}}
	\newcolumntype{L}[1]{>{\raggedright\let\newline\\\arraybackslash\hspace{0pt}}m{#1}}
	
}

\usepackage{booktabs}

\usepackage{cite}
\usepackage{amsmath,amssymb,amsfonts}
\usepackage{algorithmic}
\usepackage{graphicx}
\usepackage{textcomp}
\usepackage{xcolor}

\usepackage{lipsum,multicol}
\usepackage{enumitem}
\usepackage{algorithm}
\usepackage{algorithmic}
\usepackage{subfigure}
\usepackage[subfigure]{tocloft}
\usepackage[bottom]{footmisc}
\usepackage{multirow}
\usepackage{amsthm,amssymb,amsmath,bm}
\usepackage{rotating}
\usepackage{empheq}
\usepackage{booktabs}
\usepackage{notoccite}
\usepackage{amsmath,cases}
\usepackage{pdfpages}
\usepackage{etoolbox}
\usepackage{lastpage}
\usepackage{fancyhdr}
\usepackage{framed}
\usepackage{float}
\usepackage{subfigure}
\usepackage{comment}
\usepackage{tikz}
\usetikzlibrary{spy}
\usepackage{authblk}
\usepackage[normalem]{ulem}
\usepackage{soul}

\begin{document}

\bstctlcite{IEEEexample:BSTcontrol}
\graphicspath{{./Figures/}}

\title{Performance of Joint XR and Best Effort eMBB Traffic in 5G-Advanced Networks}


\author[1]{Pouria Paymard}
\author[2]{Abolfazl Amiri}
\author[2]{Troels E. Kolding}
\author[1, 2]{Klaus I. Pedersen}
\affil[1]{Department of Electronic Systems, Aalborg University, Aalborg, Denmark}
\affil[2]{Nokia, Aalborg, Denmark}
\affil[ ]{E-mail: pouriap@es.aau.dk}

\maketitle
    
\begin{abstract}
    In this paper, we address the joint performance of eXtended reality (XR) and best effort enhanced mobile broadband (eMBB) traffic for a 5G-Advanced system. Although XR users require stringent throughput and latency performance, operators do not lose significant additional network capacity when adding XR users to an eMBB dominated network. For instance, adding an XR service at 45 Mbps with 10 ms packet delay budget, yields close to a 45 Mbps drop in eMBB capacity. In an XR only network layer, we show how the capacity in number of supported XR users depends significantly on the rate but also the latency budget. 
    We show also how the XR service capacity is significantly reduced in the mixed service setting as the system goes into full load and other-cell interference becomes significant.  
The presented results can be used by cellular service providers to assess their networks performance of XR traffic based on their current eMBB performance, or as input to dimensioning to be able to serve certain XR traffic loads.
\end{abstract}

\section{Introduction}
    eXtended reality (XR) is one of the emerging services for 5\textsuperscript{th} generation (5G)-Advanced networks \cite{3gpp.38.838}, and is currently attracting a lot of attention in research \cite{3gpp.38.838, Opportunities_Challenges_of_XR, XR_Quality_Index, Stefano_ADRX, Performance_of_XR, JSAC_eCQI, GlobeCom_eOLLA}. Supporting XR over cellular is challenging as it calls for relatively high data rates and strict latency in order to support highly interactive real-time applications such as augmented or virtual reality (AR), virtual reality (VR) or cloud gaming (CG).   Current studies of XR over 5G have focused on a scenario where the network only carries XR traffic, see e.g. \cite{Stefano_ADRX, Performance_of_XR, JSAC_eCQI, GlobeCom_eOLLA}. For instance, various methods to optimize the user equipment (UE) power consumption for data hungry XR applications \cite{Stefano_ADRX}. Moreover, in \cite{JSAC_eCQI} and \cite{GlobeCom_eOLLA}, several link adaptation methods were studied to increase the XR capacity of 5G systems.
    
    It is clear from fundamental theory that the capacity of cellular systems will be impacted when introducing XR services. While the Shannon capacity equation is largely valid for best-effort enhanced mobile broadband (eMBB) without any quality of service (QoS) constraints, it is not applicable for XR cases with data rate, latency, and reliability constraints. For such cases, the effective bandwidth \cite{Effective_bandwidth} and effective capacity theories \cite{Effective_capacity}, as also explored in \cite{tradeoffs_RLT}, teach us that there is a price to pay in terms of lower capacity when introducing XR services with strict QoS constraints. 

    While 5G is designed to support multiplexing of diverse services such as eMBB and XR via its highly agile scheduler functionalities \cite{Klaus_Agile_5G_Scheduler}, detailed system-level performance studies of such cases are largely absent in the open literature. There are, however, a significant number of studies on the joint performance of eMBB and ultra reliable low latency communications (URLLC); see e.g. \cite{Guillermo_Joint_LA_Scheduling, Guillermo_Cost_of_URLLC}. Those studies have contributed valuable insight for cellular operators to understand the trade-offs and impact towards traditional eMBB traffic when adding URLLC services to their portfolio. URLLC traffic with typical payload sizes of 20-50 bytes and a latency constraint of 1 ms with 99.999\% reliability, is, however, significantly different from XR traffic with source data rates (SDRs) of 30-45 Mbps at 60 frames per second (fps) with 10-15 ms latency constraints at 99\% reliability. In addition, multiplexing of URLLC and eMBB traffic can benefit from preemption scheduling \cite{Preemptive_CBlayouts}, while this mechanism is not applicable for XR traffic with transmission of large payloads, typically using the same transmission time interval (TTI) sizes as eMBB.

    The objective of this study is to further study the joint performance of eMBB and XR over 5G cellular (including its evolution towards 5G-Advanced). Our objective is two-fold: (i) We want to assess how the XR capacity is impacted when eMBB best effort background traffic is added to the system, and (ii) we want to determine how a fully loaded 5G system with eMBB traffic is influenced when starting to add XR users. For the latter, we aim at assessing how much the eMBB throughput declines because of introducing XR traffic. Our goal includes a sensitivity analysis of how the network performance is impacted depending on the SDR and latency budget requirements of the XR services. We strive at presenting highly realistic system-level performance results by adopting the XR methodologies developed by the 3rd generation partnership project (3GPP) in \cite{3gpp.38.838}. Those are highly complex models that call for detailed dynamic system level simulations as it is not possible to derive analytical performance results without compromising realism.
    In Section \ref{System Model} we present the system model, followed by an outline of the evaluation methodology and performance results in Sections \ref{Evaluation Methodology} and \ref{Performance Results}. Conclusions appear in Section \ref{Conclusions}.

\section{System Model and Objectives}\label{System Model}

\subsection{Traffic Models and Key Performance Indicators (KPIs)}
    \begin{figure}[t]
		\centerline{\includegraphics[width=\linewidth,trim={0cm 0cm 0cm 0cm},clip]{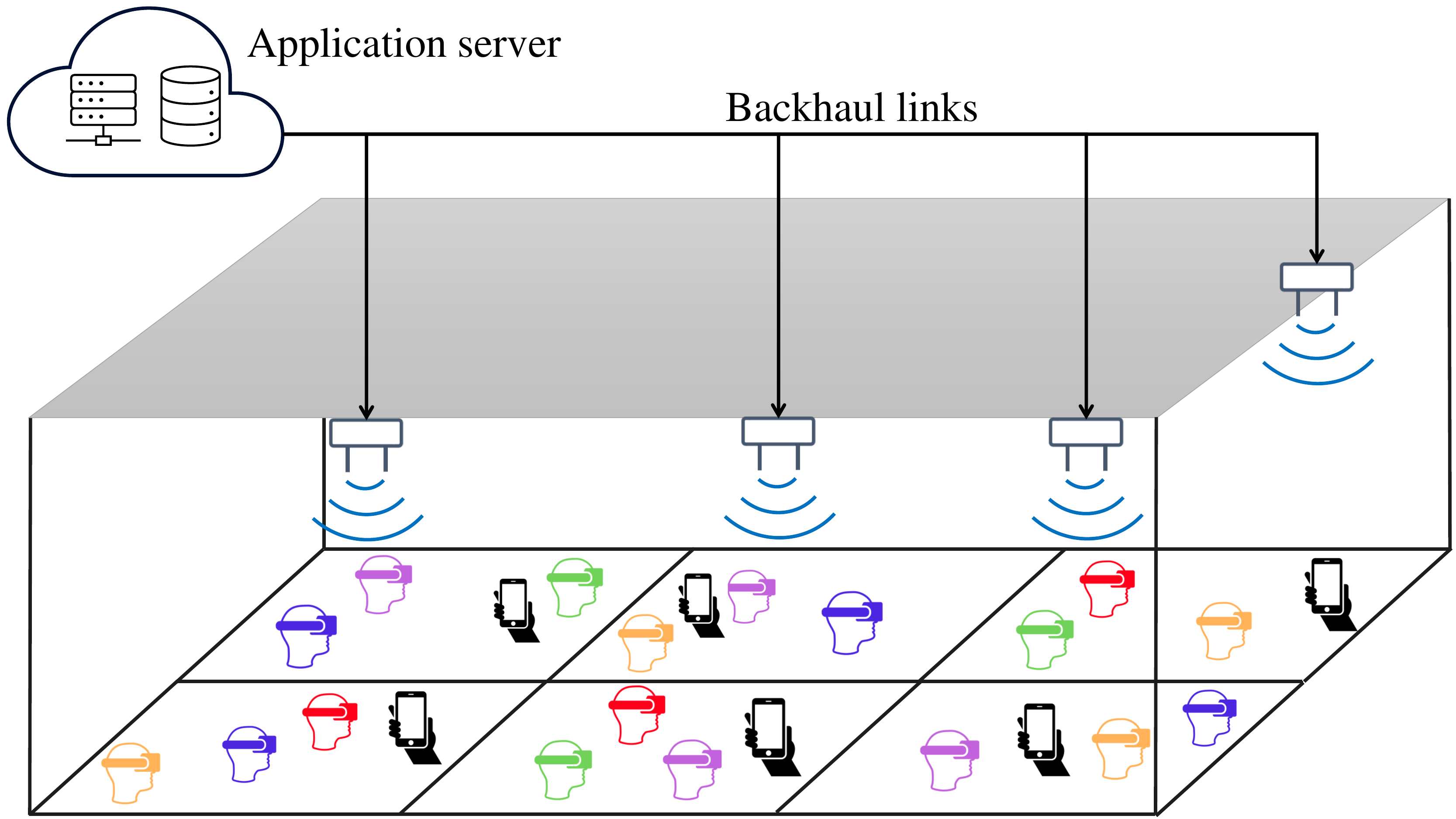}}
		\caption{Indoor hotspot deployment with one eMBB and several XR UEs per cell in the downlink scenario.}
		\label{P3:Fig:System_Model}
    \end{figure}
    Our focus is on the downlink performance and we are adopting the dynamic frame-based XR traffic models from \cite{3gpp.38.838}. XR video frames are periodically generated at the application server based on a fixed frame rate of 60 fps. The arrival time of the frames at the radio access network base station is subject to a random time-jitter that is modelled with a  truncated Gaussian distribution. The XR frames have variable sizes that are modelled with another truncated Gaussian distribution. For an XR user to be labelled as satisfied, 99\% of the XR frames shall be correctly received by the UE within the packet delay budget (PDB). The XR cell capacity is defined as the maximum number of XR users that can be supported per cell, while at least 90\% of those are satisfied. In this study, we consider XR cases with an average SDR of 30 Mbps and 45 Mbps, as well as PDB values in the range of 5-30 ms.
    According to \cite{3gpp.38.838}, the case with 10 ms PDB and 30 Mbps (45 Mbps) corresponds to single-eye (dual-eye) VR traffic with 4K video quality, while the case with 15 ms PDB corresponds to CG applications.

    For the sake of simplicity, we model the best effort eMBB traffic according to the full buffer model. That means, there is always buffered data at the base station to be transmitted to the eMBB UE. The main KPI for the eMBB traffic is the average experienced throughput. Note that for cases with eMBB traffic in a cell, the cell would constantly operate at full load as having sufficient pending data at the base station to transmit on all available radio resources. For cases with only XR users, the system would be operating at fractional load conditions, where not all radio resources would be in use, as sometimes a cell have no pending data to transmit, or not enough data to transmit on all resources.

\subsection{Radio Access Network}
\begin{figure}[t]
		\centerline{\includegraphics[width=\linewidth,trim={0cm 0cm 0cm 0cm},clip]{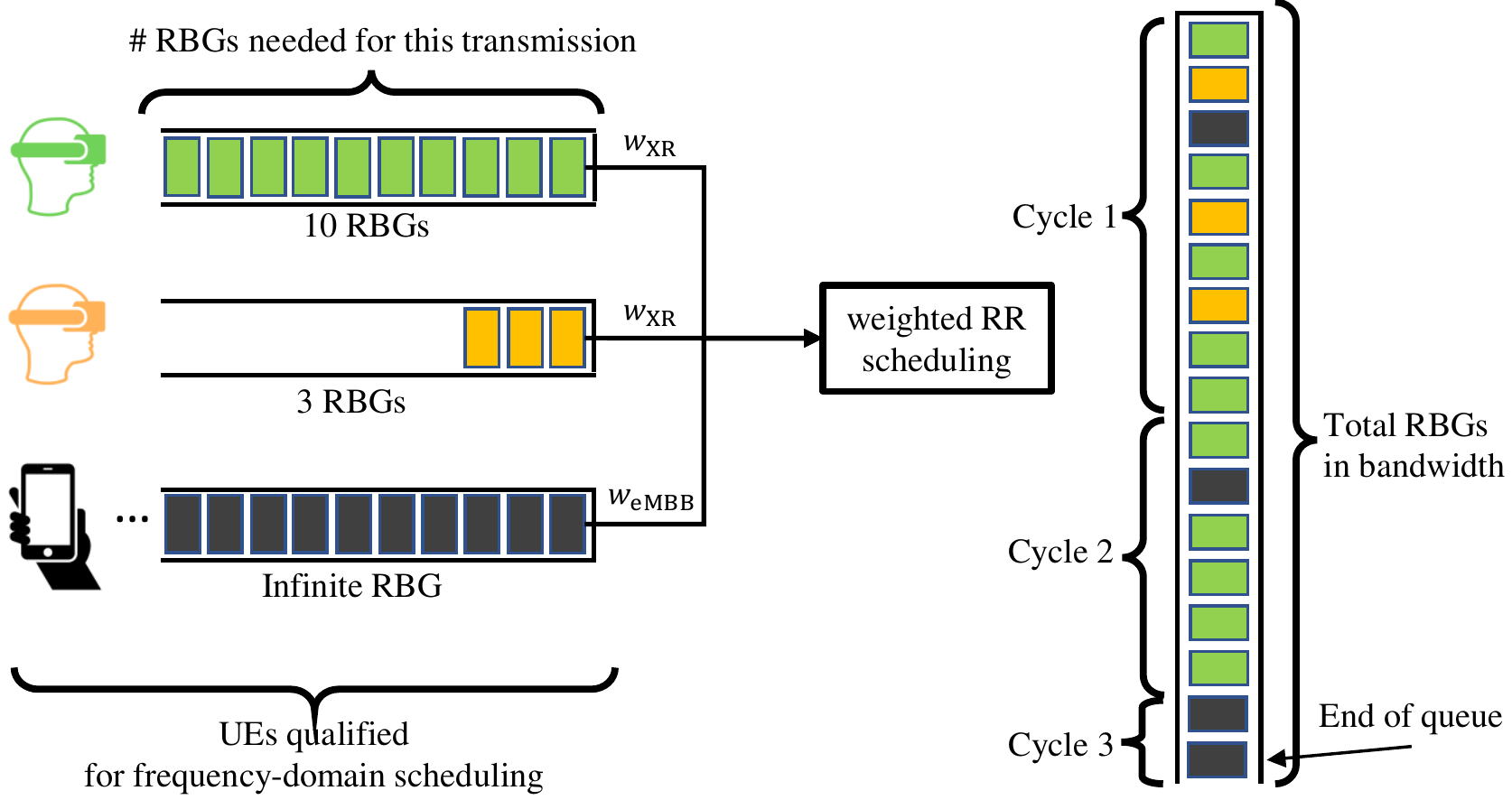}}
		\caption{Example of interleaved WRR scheduling for two different user groups, i.e., eMBB and XR, assuming that $w_{\text{eMBB}}=1$ and $w_{\text{XR}}=5$.}
		\label{P2:Fig:Scheduler}
    \end{figure}

    The system model comprises a 5G single-frequency layer network with multiple base stations and UEs in line with 3GPP New Radio scenarios in \cite{3gpp.38.838}. We adopt the indoor hotspot scenario \cite{3gpp.38.901} as illustrated in Fig. \ref{P3:Fig:System_Model}. Deployment at 4 GHz with an unpaired 100 MHz carrier, assuming 30 kHz subcarrier spacing (SCS). Dynamic scheduling at every TTI corresponding to one slot of 14 OFDM symbols (0.5 ms) is assumed. Dynamic link adaptation (LA) with adaptive modulation coding scheme (MCS) selection based on channel quality indicator (CQI) feedback is assumed for each scheduled user. This is assuming the use of code block group (CBG)-based transmissions as used also in \cite{GlobeCom_eOLLA} and \cite{JSAC_eCQI}. For cases where the UE fails to correctly decode all the scheduled CBGs, the base station will subsequently only schedule retransmission of the failed CBGs, and the UE would apply Chase soft combining. Thus, asynchronous CBG-based Hybrid Automatic Repeat reQuest (HARQ) is used. The scheduling policy at each base station is as follows: We assume that scheduling of pending HARQ retransmissions is first prioritized. Secondly, the scheduling of the first transmission XR data is prioritized, followed by eMBB data. This is realized with a weighted round robin (WRR) policy as described in \cite{Weighted_Round_Robin}. Fig. \ref{P2:Fig:Scheduler} shows an example of how resource block groups (RBG) are allocated to XR and eMBB UEs when using WRR scheduling policy as per UEs' weights.
    
\section{Evaluation Methodology}\label{Evaluation Methodology}
    \begin{table}[t]
		\centering
		\caption{Summary of System-level Evaluation Parameters}
		\begin{tabular}{c c}
			\textbf{Parameter} & \textbf{Setting} \\  \toprule 
			Simulation time & 6 sec per run \\ 
			Simulation runs  & 5 runs \\ \hline
			Deployment & Indoor Hotspot (120m$\times$50m) \\ 
			Layout  & 12 cells  \\ 
			Inter-site Distance & 20 m \\ \hline
			TDD Frame structure & DDDSU \\ 
			TTI length & 14 OFDM symbols \\
			PDCCH overhead & 1 OFDM symbol per D and S slot \\
			Carrier frequency & 4 GHz \\ 
			System Bandwidth & 100 MHz \\ 
			SCS & 30 kHz \\ \hline
			MIMO scheme & SU-MIMO with rank 1  \\ 
			Modulation & QPSK to 256QAM \\
			gNB height & 3 m  \\ 
			gNB Tx power & 31 dBm \\ 
			gNB Tx processing delay & 2.75 OFDM symbols \\ 
			\multirow{2}*{gNB antenna} & 1 panel with 32 elements \\
			& (4 × 4 and 2 polarization) \\ \hline
			UE speed & 3 km/h  \\ 
			UE height & 1.5 m  \\ 
			UE Rx processing delay & 6 OFDM symbols \\
			UE antenna & 2 dual-polarized antennas \\ \hline
			Jitter distribution & $\mathcal{TN}(0, 2, -4, 4)$ ms \\
			XR frame size (30Mbps) & $\mathcal{TN}(62.5, 6.25, 31.25, 93.75)$ kB \\
			XR frame size (45Mbps) & $\mathcal{TN}(93.75, 9.8, 46.875, 140.625)$ kB \\
			XR frame rate & 60 fps \\
			XR SDR & 30 Mbps, 45 Mbps \\ \hline
			Scheduler & WRR \\ 
			$w_{\text{XR}}$ & 20 \\ 
			$w_{\text{eMBB}}$ &1 \\ \hline
			HARQ scheme & CBG-based HARQ retransmissions \\ 
            HARQ combining method & Chase soft combining \\
			CQI & Periodic CQI every 2ms  \\
			Target CBG error rate & 2 failed out 8 CBGs \\ \toprule
		\end{tabular}
		\label{P3:Table:System_Parameters}
	\end{table} 
    The evaluation methodology and KPIs to assess the network performance is aligned with the 3GPP technical report in \cite{3gpp.38.838}, as also adopted in the recent publications \cite{petrov2022extended, GlobeCom_eOLLA}.
    A summary of the simulation parameters is captured in Table \ref{P3:Table:System_Parameters}. Note that $\mathcal{TN}$(mean, standard deviation, min value, max value) denote the truncated Gaussian distribution. The system-level evaluation methodology include the major  radio access network mechanisms. Among others, it includes the dynamic packet scheduling, CQI measurements and reporting, outer loop LA (OLLA) algorithm, HARQ mechanism, and time- and frequency-varying inter-cell interference.
    Users are spatially uniformly distributed, with an equal number of UEs per cell.
    The used 3GPP 3D indoor hotspot radio channel propagation model is calibrated against alike results published in 3GPP. The radio frame configuration is the same, and fully synchronized, for all cells in the system. It follows a DDDSU slot repetition pattern with three downlink, one special, and one uplink slot format. The special slot includes 10 downlink symbols, two symbols for the guard period, and two symbols for uplink. Physical downlink control channel (PDCCH) transmissions appear in the first downlink symbol of slots where this is present.

    For each scheduled transmission, the signal-to-interference-noise-ratio (SINR) per resource element (RE) at the receiver end is calculated. The SINR calculation takes the effect of the single-user multiple-input multiple-output (SU-MIMO) closed loop scheme into account with the used transmitter precoder and minimum mean square error interference rejection combining (MMSE-IRC) receiver. Note that we assume simple rank one transmissions only in this study. Following this, the per-RE SINR values for the transmission are mapped to an effective SINR per CBG and TB and, subsequently, the mean mutual information per bit (MMIB) is computed \cite{tang2010mean}. Given the effective SINR, MMIB, and the used MCS for the transmission, the BLEP of the transmission is obtained from a look-up table. That look-up table is obtained from extensive link-level simulations. A failure to correctly decode a transmission will trigger an asynchronous adaptive HARQ retransmission. We set the maximum number of HARQ retransmissions to 3. In line with our system model in Section \ref{System Model}, we assume CBG-based HARQ operation.  
    
    Each simulation is run for a time corresponding to the transmission of at least 360 XR packets for each call.
    Considering uncorrelated samples, this allows us to estimate if 99\% of the packets are correctly received at the UE with an error margin of at most $\pm 0.256\%$ at a confidence level of 99\% as is the criteria for labelling a user as satisfied. For each simulation with $N$ XR users per cell, and 12 cells, we collect statistics from $12 \times N$ calls, with typical values of $N$ being in the range of 3 to 6. We repeat such simulations $M=5$ times (aka $M$ simulation runs), so we have statistics from $12 \times N \times M$ XR calls. With $M=5$ and $N=5$, this gives us statistics from 300 XR calls, which is sufficient to estimate if 90\% of the XR users are satisfied as per the XR capacity definition metric. Using these settings, we simulate 12000 TTIs for each of the 12 cells, repeating it $M=5$ times. This is sufficient for also getting reliable statistics for eMBB average cell throughput by monitoring the throughput samples per TTI.

\section{Performance Results}\label{Performance Results}
\subsection{Impact on the XR performance from eMBB}
    We first investigate how the XR performance is impacted when eMBB traffic is present in the network. From an isolated single-cell perspective, there would in principle be no impact from adding eMBB traffic to the network. That is because our scheduler prioritizes XR traffic over eMBB and there is no intra-cell cross-link interference between users.
    However, in a multi-cell network, adding full buffer eMBB traffic on top of XR means that each cell will operate at 100\% radio resource utilization.
    \begin{figure}[t]
		\centerline{\includegraphics[width=0.98\linewidth,trim={0cm 0cm 0cm 0cm},clip]{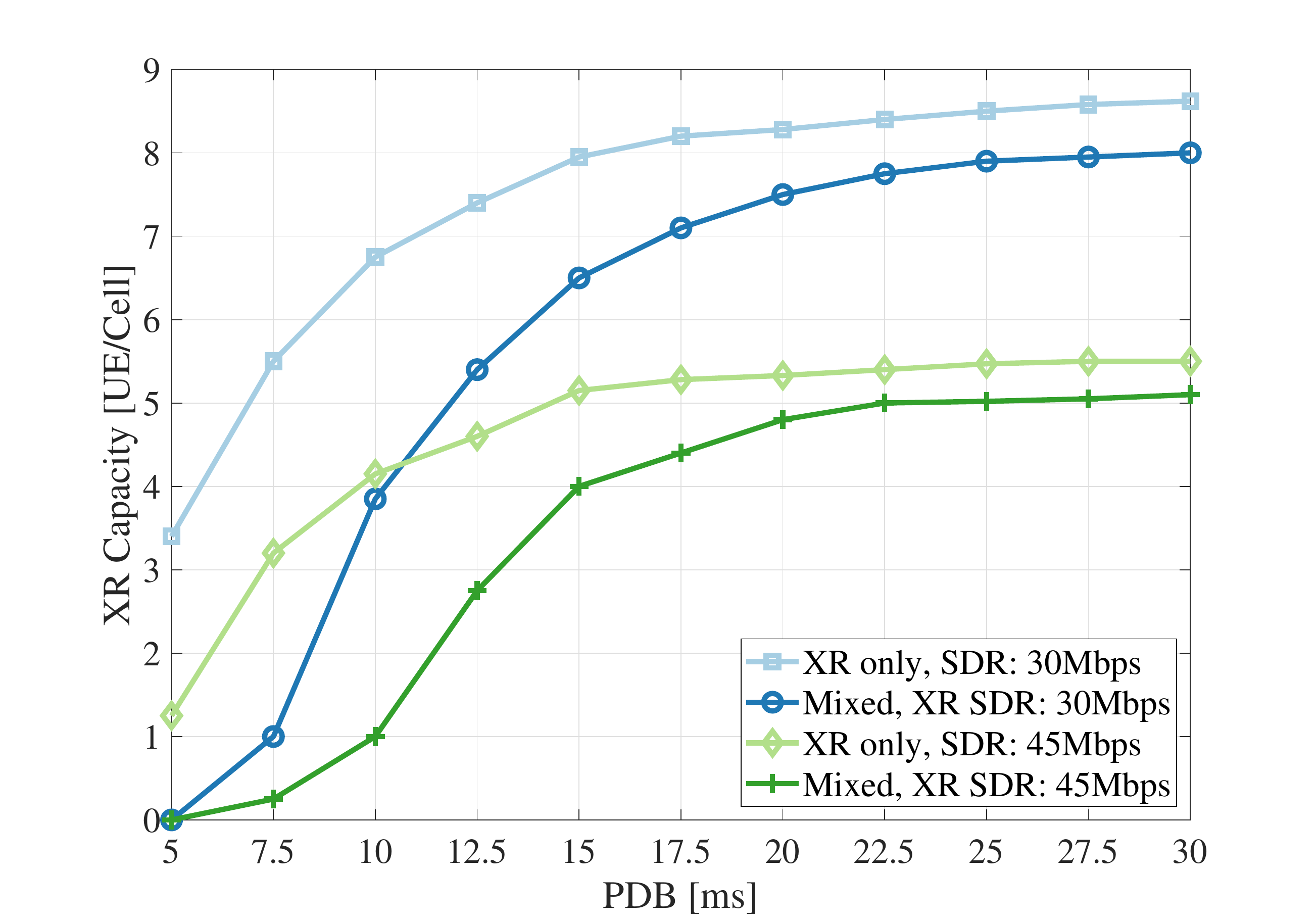}}
		\caption{XR capacity versus PDB for different XR SDRs with/without eMBB traffic.}
		\label{P3:Fig:Capacity_vs_PDB}
    \end{figure}
    \begin{figure}[t]
		\centerline{\includegraphics[width=0.98\linewidth,trim={0cm 0cm 0cm 0cm},clip]{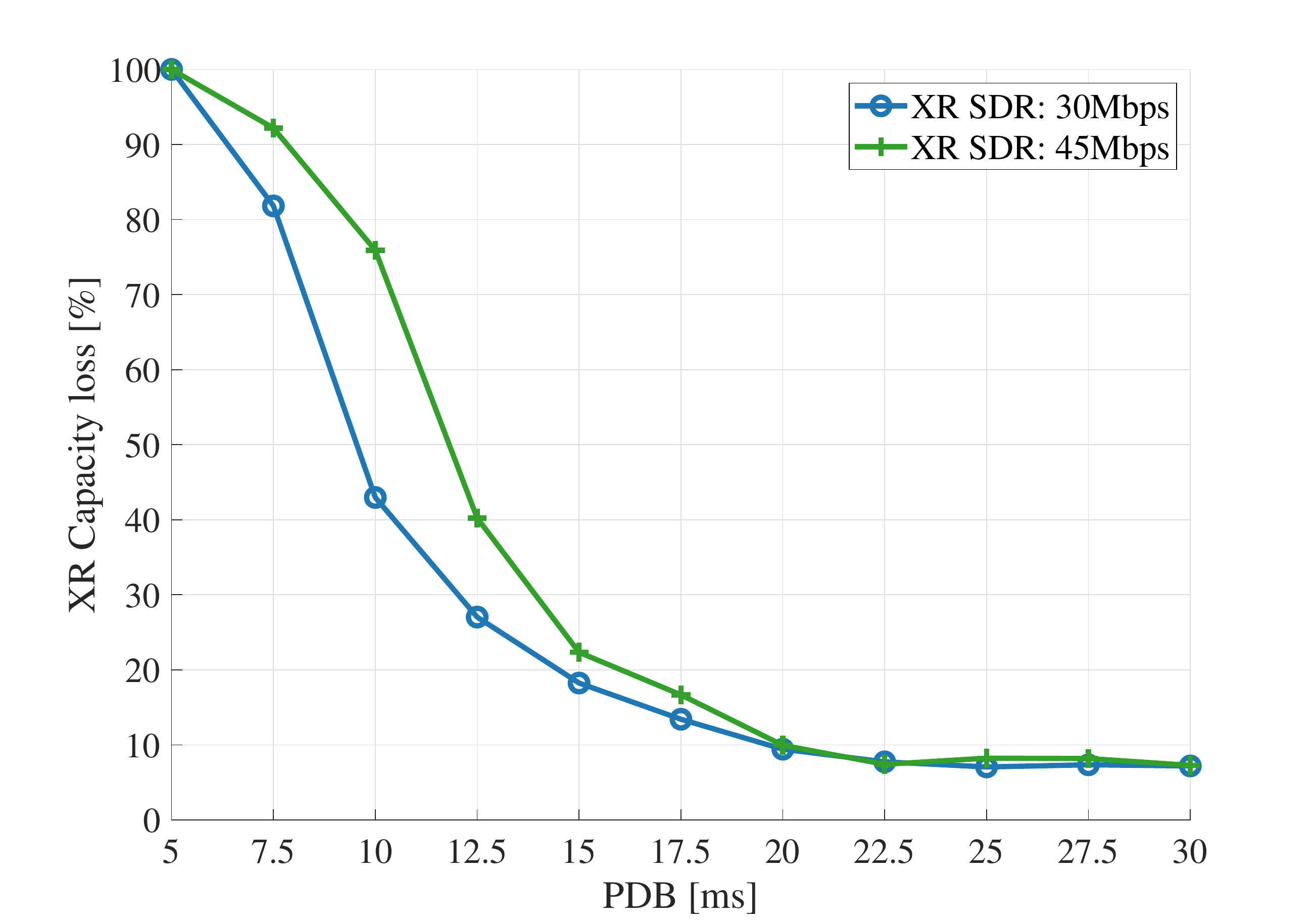}}
		\caption{XR capacity loss versus PDB for different XR SDRs.}
		\label{P3:Fig:Capacity_Loss_vs_PDB}
    \end{figure}
    Fig. \ref{P3:Fig:Capacity_vs_PDB} shows the XR capacity with/without (w/o) eMBB best effort traffic, while XR capacity loss for different XR SDRs and PDB values is illustrated in Fig. \ref{P3:Fig:Capacity_Loss_vs_PDB}. 
    Recall that the XR capacity is defined as the maximum number of supported UEs per cell, while at least 90\% of those are satisfied (each satisfied UE must receive more than 99\% of its packets within the PDB).
    The XR capacity naturally declines for higher SDR values, while it increases if the PDB is relaxed. The XR capacity loss is most significant for high data rate XR cases with tight PDB requirements, while it is less if the PDB is relaxed. At 45 Mbps SDR, the XR capacity loss equals 75\% and 10\% for PDB values of 10ms and 20ms, respectively. This is because less queuing delays are tolerated for the stricter XR QoS requirements. As XR traffic is always prioritized over eMBB traffic, the queuing delays occur mainly due increased other-cell interference caused by the eMBB users. 
    
\begin{figure}[t]
		\centerline{\includegraphics[width=0.98\linewidth,trim={0cm 0cm 0cm 0cm},clip]{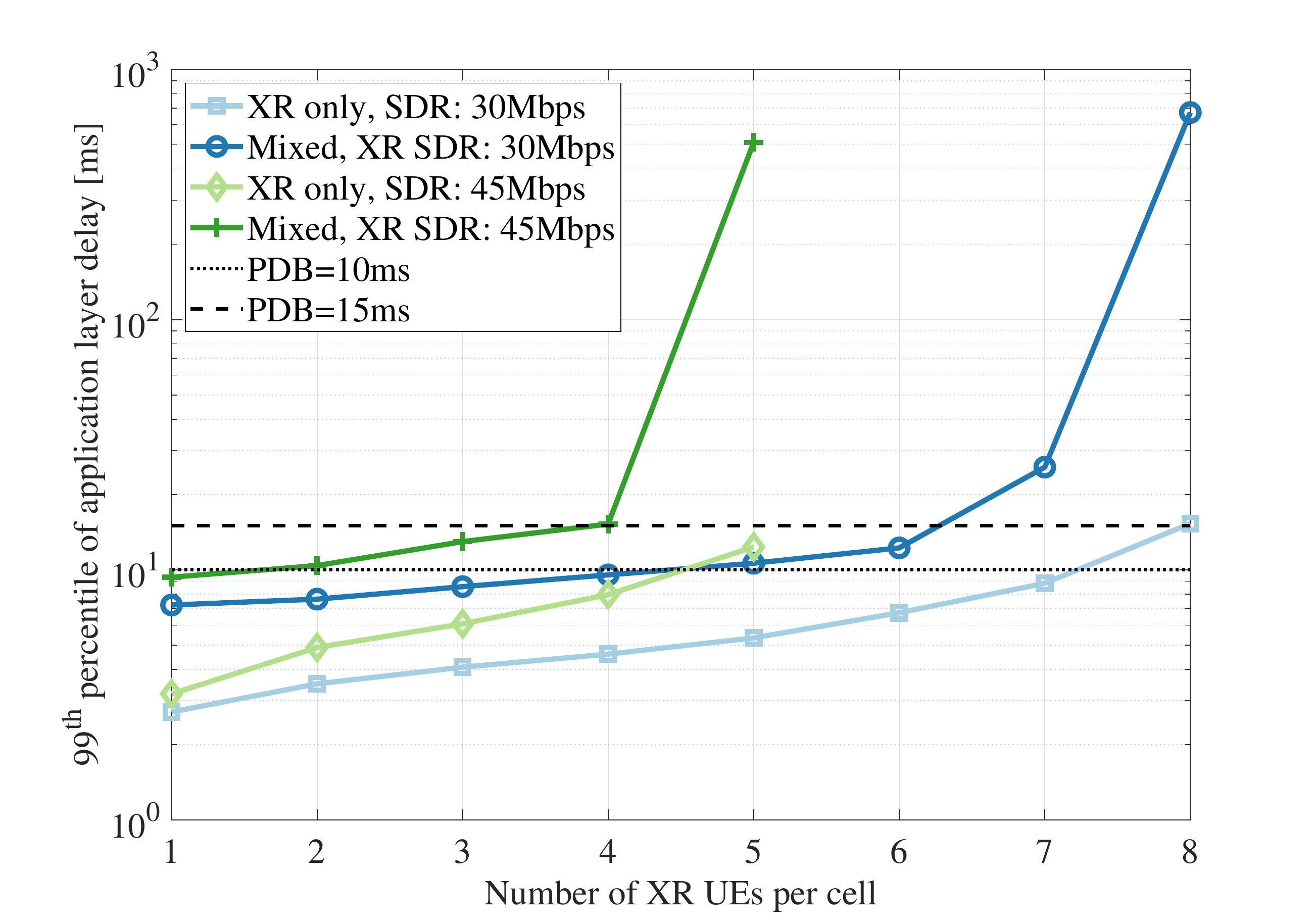}}
		\caption{99\textsuperscript{th} percentile of eCDF of application layer delay versus the number of XR UEs per cell for cases with/without eMBB.}
		\label{P2:Fig:Delay_99th}
    \end{figure}

    The 99-percentile of the empirical cumulative distribution function (eCDF) of the application layer packet latency is plotted in Fig. \ref{P2:Fig:Delay_99th} for two XR SDR values. As can be seen for the cases with mixed eMBB and XR traffic, the latency grows faster. This is because queuing is more likely to happen when a full-buffer eMBB traffic is present and leads to more interference, and hence lower air interface throughput performance. As an example for 30 Mbps SDR, the experienced delay for the case with one XR and one eMBB UE is higher than that of having 6 XR UEs in XR only scenario.
    
     \begin{figure}[t]
		\centerline{\includegraphics[width=0.98\linewidth,trim={0cm 0cm 0cm 0cm},clip]{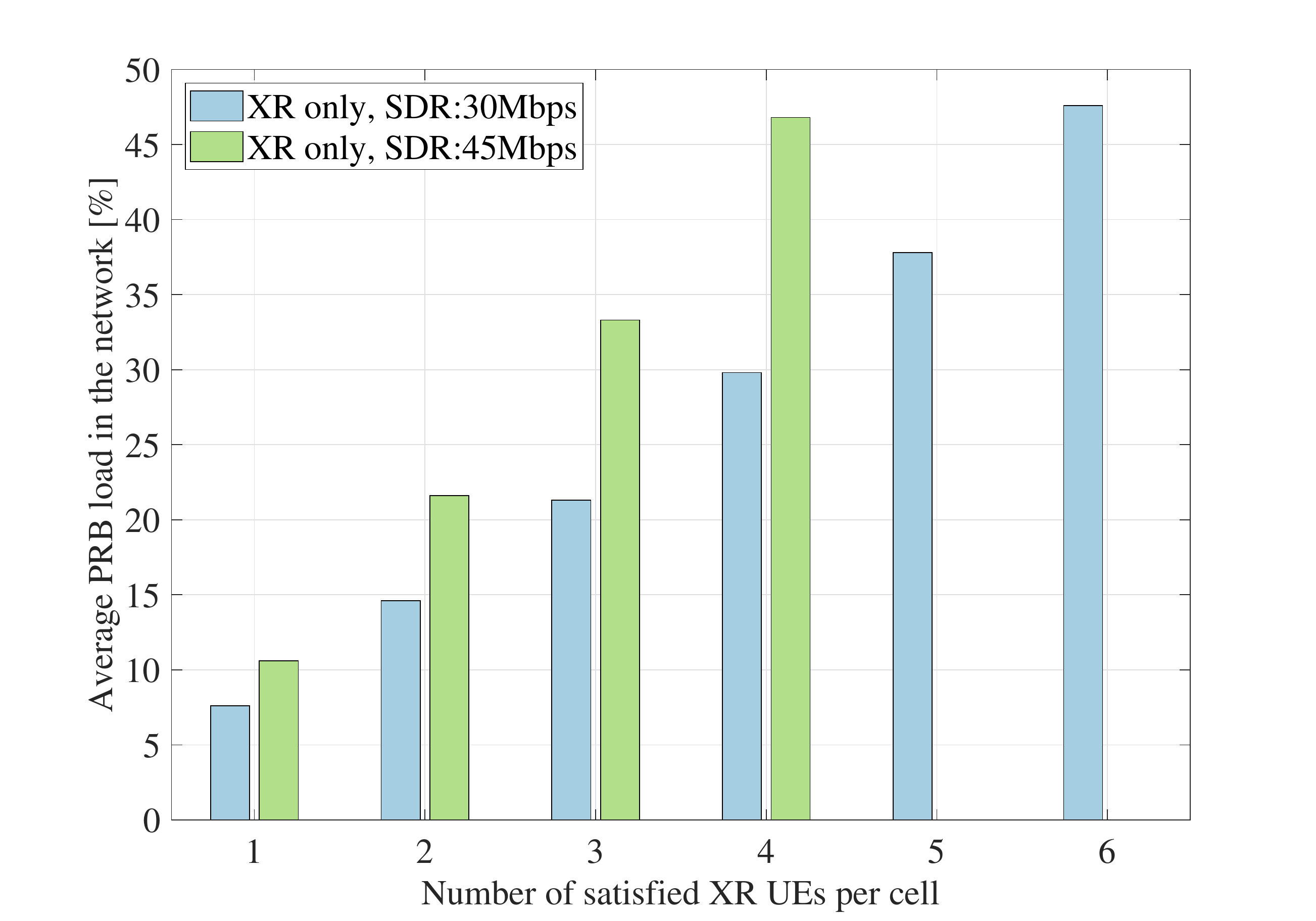}}
		\caption{Average PRB utilization versus the number of XR UEs per cell for different XR SDRs.}
		\label{P2:Fig:PRB_Load_Bars}
    \end{figure}
    
    \begin{figure}[t]
		\centerline{\includegraphics[width=0.98\linewidth,trim={0cm 0cm 0cm 0cm},clip]{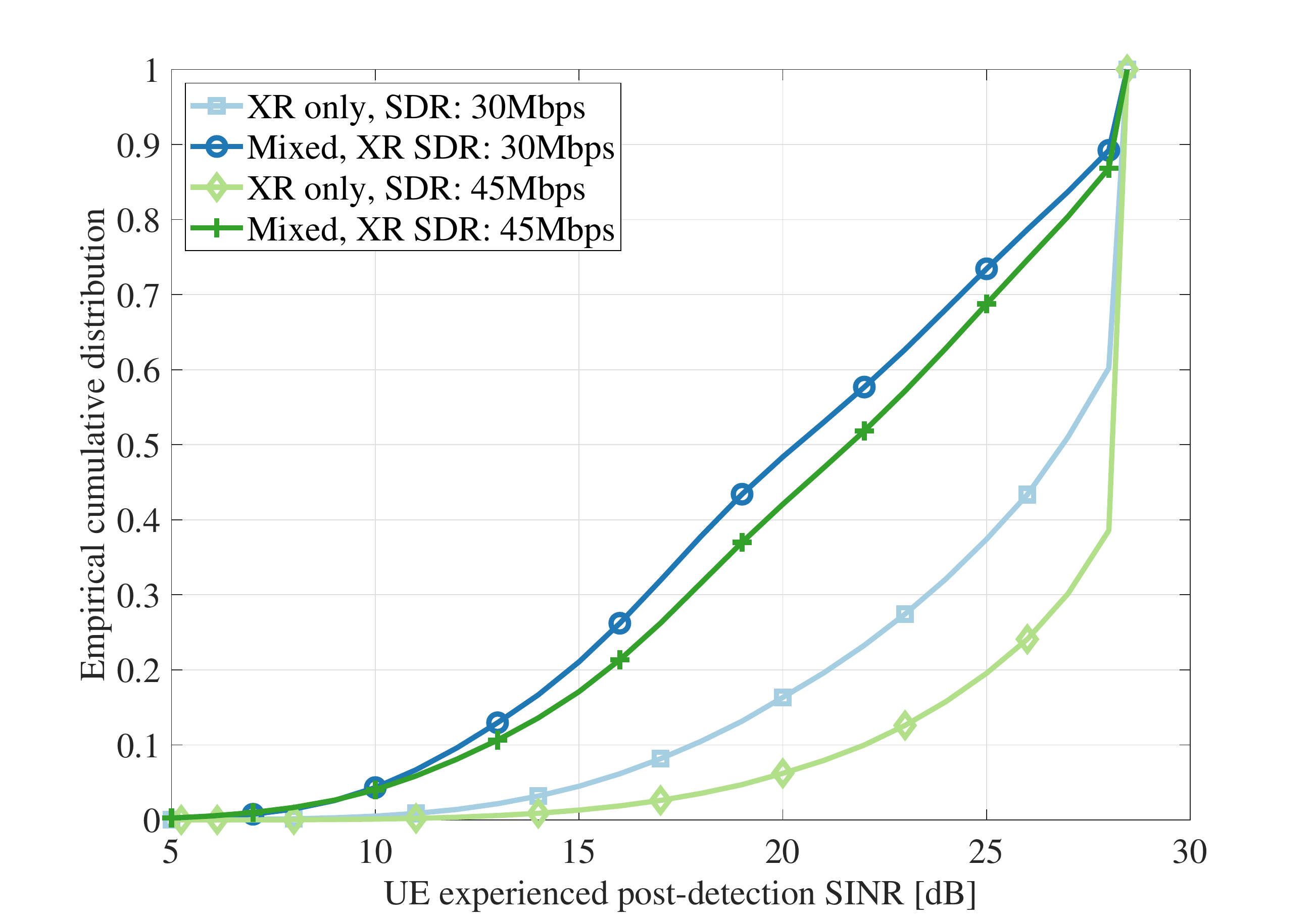}}
		\caption{eCDF of the UE experienced post-detection SINR for cases with/without eMBB and the maximum XR capacity, i.e., 7 UE/cell for 30Mbps and 4 UE/cell for 45Mbps.}
		\label{P2:Fig:SINR}
    \end{figure}

Fig. \ref{P2:Fig:PRB_Load_Bars} shows the physical resource block (PRB) utilization for different XR loads without eMBB traffic. As expected, it is often far below full load conditions for XR-only scenarios, while with eMBB traffic added, it reaches full load, i.e. 100\% PRB utilization, and therefore increased interference. Both cases seem to arrive at maximum XR capacity when the PRB utilization is around 45-50\% leading to medium othercell interference.

The eCDF of the UE experienced post-detection SINR is plotted in Fig. \ref{P2:Fig:SINR}. Results are plotted for cases w/o eMBB for the maximum XR capacity of 4 users (for 45 Mbps with 10 ms PDB) and 7 users (for 30 Mbps with 15ms PDB). Note that this capacity applies for the XR-only case, so for the mixed results those XR users do not meet their QoS requirement. 
As expected, it is observed that the SINR distribution is shifted to the lower values after adding eMBB traffic. In fact, at the median level, it decreases by 4.7 to 5.6 dB. The highest SINR decline is observed for the case with 4 XR users at 45 Mbps and 10 ms PDB, where the radio resource utilization is only 47\% (see Fig. \ref{P2:Fig:PRB_Load_Bars}) when there is no eMBB. For the case with 7 XR users at 30 Mbps and 15 ms PDB, the radio resource utilization is 60\% w/o eMBB, and hence the consequence of adding eMBB (and reaching 100\% radio resource utilization) is a more modest SINR decline of 4.7 dB.

\subsection{Impact on the eMBB performance from XR}

    \begin{figure}[t]
		\centerline{\includegraphics[width=0.98\linewidth,trim={0cm 0cm 0cm 0cm},clip]{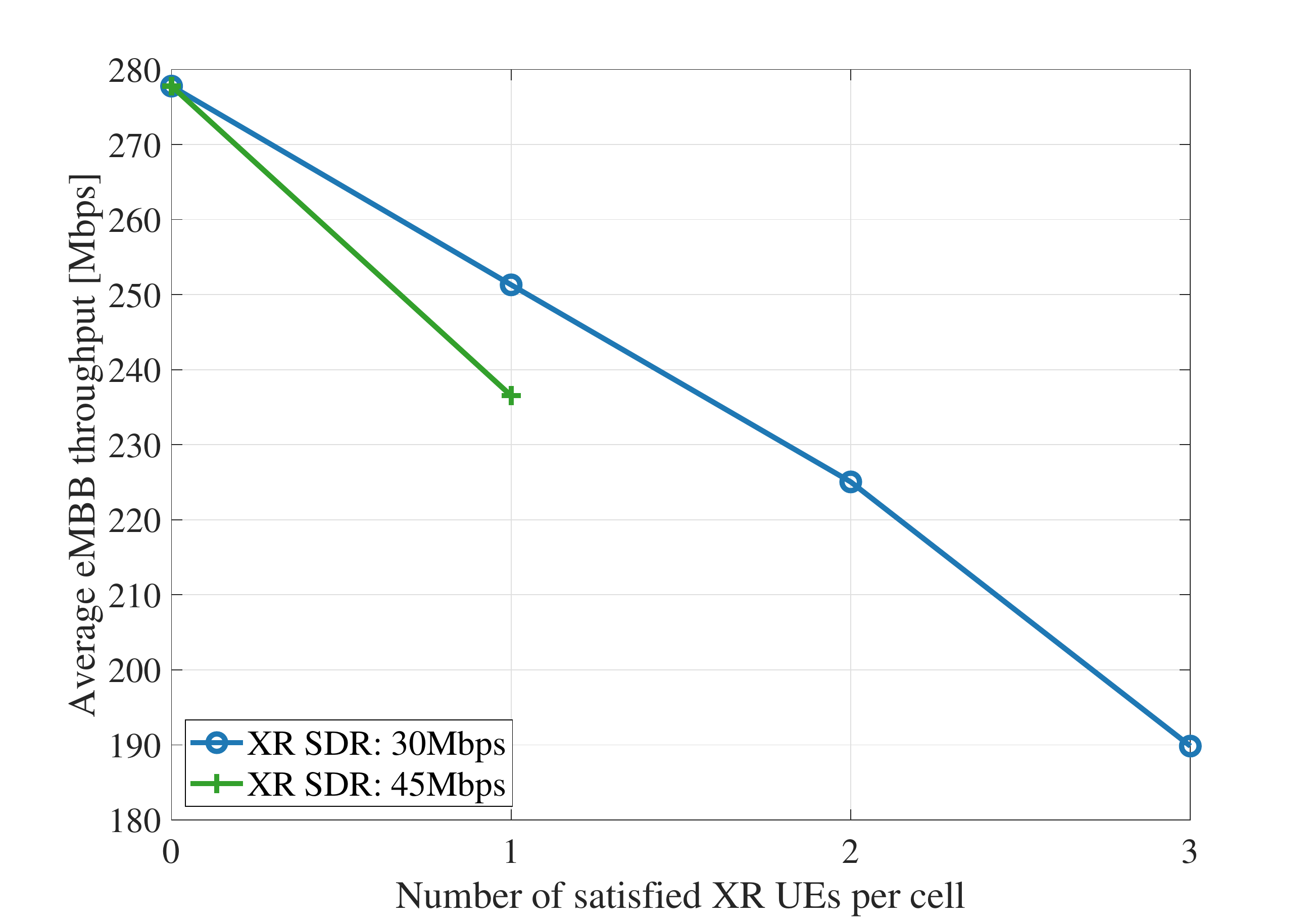}}
		\caption{Average eMBB throughput versus the number of XR UEs per cell when XR SDR is either 30Mbps or 45Mbps.}
		\label{P3:Fig:eMBB_Throughput_Avg}
    \end{figure}
    
    \begin{figure}[t]
		\centerline{\includegraphics[width=0.98\linewidth,trim={0cm 0cm 0cm 0cm},clip]{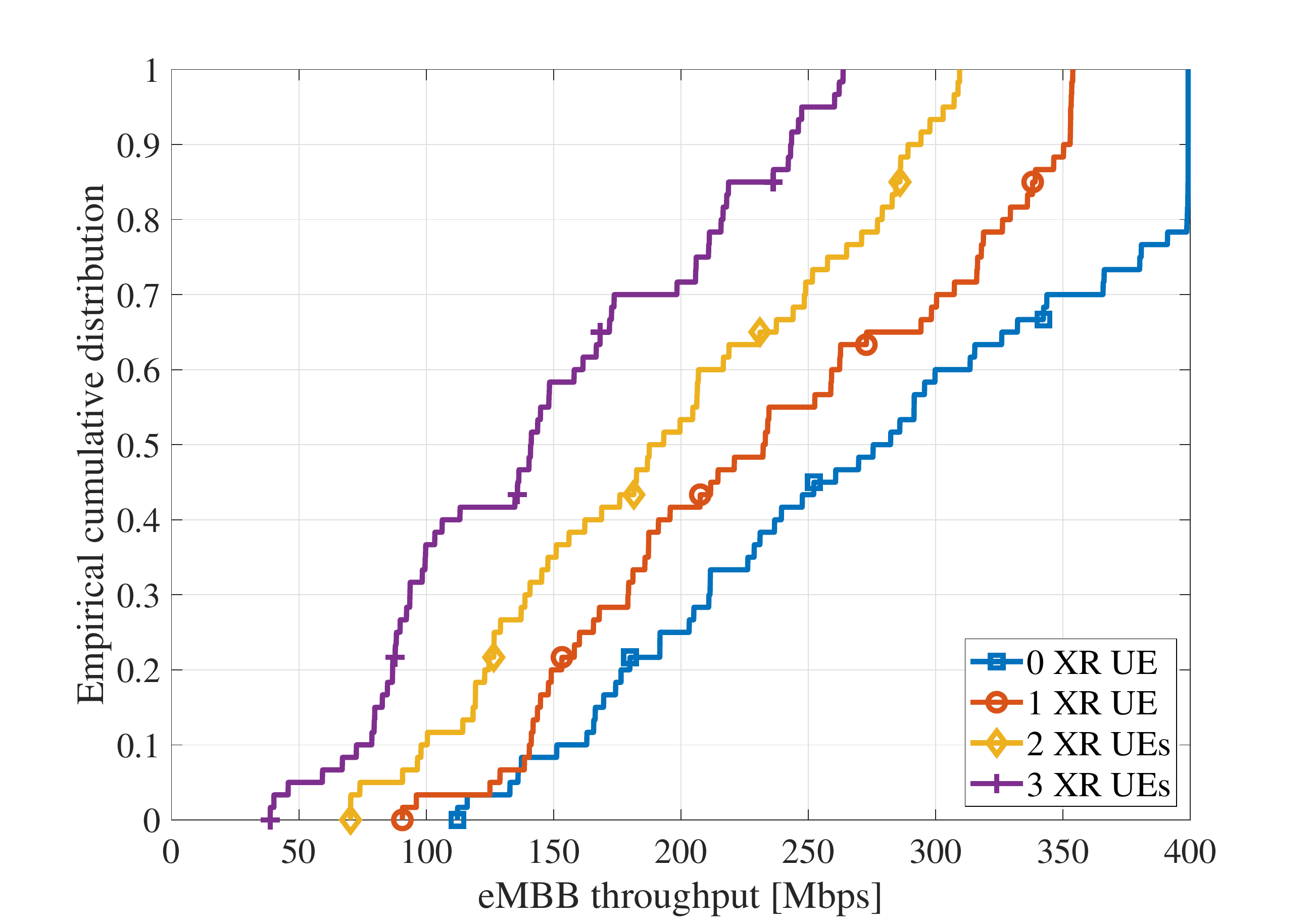}}
		\caption{eCDF of eMBB throughput when XR SDR is 30Mbps.}
		\label{P3:Fig:eMBB_Throughput_CDF}
    \end{figure}
    
    In many deployment cases, XR users will be added to a network that service primarily eMBB users and we therefore next consider the impact to eMBB capacity by adding XR service. Fig. \ref{P3:Fig:eMBB_Throughput_Avg} shows the average eMBB cell throughput as a function of added XR users with either 30 Mbps or 45 Mbps SDRs. Furthermore, the eCDF of the eMBB cell throughput in Fig. \ref{P3:Fig:eMBB_Throughput_CDF} for the case with 45Mbps XR SDR.
    With no XR UEs, the average cell throughput equals 279 Mbps. 
    As three 30 Mbps XR users are added, the eMBB throughput drops by roughly 90 Mbps, so there is 1-1 trade between XR and eMBB capacity. The same is true when adding a single 45 Mbps user, the eMBB throughput drops by approximately 45 Mbps as well. Thus, the spectral efficiency of the mixed service network is not reduced adding XR services. This is in contrast to for instance     
URLLC traffic which was studied in \cite{Guillermo_Cost_of_URLLC} when adding URLLC services significantly reduce the remaining eMBB capacity. 

\section{Conclusions}\label{Conclusions}
    In this paper, we have studied the performance of a 5G system with mixed XR and eMBB traffic, where the XR traffic has strict QoS requirements, while the eMBB is assumed to be best effort.
    We have presented dynamic system-level simulation results in line with the latest industry-agreed assumptions. As XR users are added to a scenario with existing eMBB users, there is a 1-1 trade-off. Ex. the Mbps of XR traffic added yields a similar loss of eMBB traffic. Adding eMBB traffic to an XR-only network, our results show significant degradation of XR capacity by 75\% for a strict PDB of 10ms, while the decline is reduced to 10\% for a more relaxed PDB of 20 ms. The decline in XR capacity is mainly contributed to the increased other-cell interference in the system.
    The presented findings, can serve as input to help cellular service providers on how to dimension their network when introducing XR traffic.

\hyphenation{op-tical net-works semi-conduc-tor}
\bibliographystyle{IEEEtran}

\end{document}